# Thermal annealing enhancement of Josephson critical currents in ferromagnetic CoFeB

Sachio Komori*, Juliet E. Thompson*

Guang Yang, Graham Kimbell, Nadia Stelmashenko

Mark G. Blamire and Jason W. A. Robinson†

Department of Materials Science & Metallurgy, University of Cambridge,

27 Charles Babbage Road, Cambridge CB3 0FS, United Kingdom

**The electrical and structural properties of $Co_{40}Fe_{40}B_{20}$ (CoFeB) alloy are tunable with thermal annealing. This is key in the optimization of CoFeB-based spintronic devices, where the advantageously low magnetic coercivity, high spin polarization, and controllable magnetocrystalline anisotropy are utilised. So far, there has been no report on superconducting devices based on CoFeB. Here, we report Nb/CoFeB/Nb Josephson devices and demonstrate an enhancement of the critical current by up to 700% following thermal annealing due to increased structural ordering of the CoFeB. The results demonstrate that CoFeB is a promising material for the development of superconducting spintronic devices.**

*These authors contributed equally to this work.

†Corresponding author: jjr33@cam.ac.uk

Following the discovery of giant magnetoresistance [1,2] and the development of the first spin-valves [3,4], $Co_{40}Fe_{40}B_{20}$ (CoFeB) was identified as an alternative magnetic material to those that had previously been employed, particularly due to its low magnetic anisotropy and low switching energy [5,6]. In CoFeB spintronic devices, the magnetoresistance can be optimized through a thermal annealing-induced structural transition from amorphous to crystalline [7]. Specifically, for CoFeB/MgO/CoFeB magnetic tunnel junctions, the crystallization of CoFeB leads to high tunneling magnetoresistance, exceeding 600% at room temperature [8]; neither CoFeB nor CoFe as-grown devices display comparable efficiencies. Studies of diffusive (i.e. without a tunnel junction) CoFeB spin-valves have also demonstrated larger giant magnetoresistance effects following annealing treatment [9].

Although the advantageous properties of CoFeB-based spintronic devices and their controllability through annealing have been well recognized in the field of spintronics, there has been no report on superconducting spintronic devices involving CoFeB. This may be partly due to the strong magnetic exchange energy and the high resistivity of CoFeB [10], which should strongly quench the superconducting proximity effect, making it challenging to investigate the coupling of superconductivity and magnetism. Here, we report Nb/CoFeB/Nb Josephson devices with thin (< 5 nm) CoFeB barriers and investigate the effect of thermal annealing on the critical current ($I_c$). From measurement of the Josephson critical currents versus CoFeB barrier thickness, we determine a superconducting coherence length in CoFeB of approximately 2 nm. Annealing the devices at 400°C for 30 minutes in vacuum results in the increase in the critical current by as much as 700% for a CoFeB thickness of 4 nm. We associate this enhancement of the Josephson current with an increase in the electron mean free path lengths for charge and spin-flip scatter in CoFeB along with improved transparency at the Nb/CoFeB interfaces.

Nb(300 nm)/CoFeB($d_{CoFeB}$)/ Nb(300 nm) trilayer stacks were fabricated on 5 mm × 5 mm quartz substrates by dc magnetron sputtering in an ultrahigh-vacuum chamber with a base pressure better than $10^{-6}$ Pa. The sputtering targets ($Co_{40}Fe_{40}B_{20}$, Nb) were pre-sputtered for 20 minutes to clean their surfaces. The films were grown in Ar at a pressure of 1.5 Pa at room temperature. Multiple quartz substrates were placed on a rotating circular table that passed below a series of stationary magnetrons. A series of stacks were prepared with different CoFeB thicknesses ($d_{CoFeB}$ = 1.5 – 4.5 nm) between 300-nm-thick layers of Nb in a single deposition. Layer thicknesses were controlled by adjusting the angular speed of the rotating table.

Current-perpendicular-to-plane Nb(300 nm)/CoFeB($d_{CoFeB}$)/Nb(300 nm) nanopillar devices with square cross-sectional areas of approximately $A = 500 \times 500$ nm$^2$ were fabricated using a focused beam of Ga ions as described elsewhere [11]. A pulse-tube cryogen-free measurement system (Cryogenic Ltd) was used to cool the devices down to 1.6 K. Resistivity and current-voltage $I(V)$ characteristics of the nanopillars were measured in a four-point configuration using the differential conductance mode of a Keithley 6221 AC-current source and a 2182A nanovoltmeter. The Josephson critical current ($I_c$) and normal state resistance ($R_N$) of each device were determined by fitting the $I(V)$ characteristics to the resistively shunted junction model $V = R_N(I^2-I_c^2)^{0.5}$. $I_c$ was modulated by applying a magnetic field ($H$) parallel to the plane and perpendicular to the current direction. Electrical measurements were performed on nanopillars both before and after thermal annealing. Thermal annealing was performed at 400°C for 30 minutes in vacuum ($10^{-5}$ Pa) – the typical post-anneal condition [8,12] to promote crystallisation. Annealed CoFeB has lower resistivity than amorphous CoFeB deposited at room temperature [13].

A typical $I_c(H)$ Fraunhofer pattern for a Nb(300 nm)/CoFeB(3.5 nm)/ Nb(300 nm) nanopillar at 1.6 K before (solid curves) and after (dashed curves) thermal annealing is shown in Fig. 1(a) [see supplementary Fig. S1 for all the $I_c(H)$ data recorded in this study]. $I_c(H)$ is hysteretic and the maximum values of $I_c$ are obtained at non-zero applied fields ($\mu_0 H = \delta$) due to the intrinsic barrier magnetization [14,15]. In Fig. 1(b), we have plotted $\delta$ at 1.6 K versus $d_{CoFeB}$, which shows a linear increase in $\delta$ with $d_{CoFeB}$. By fitting $\delta$ versus $d_{CoFeB}$ to $\delta = M_s(d_{CoFeB} - d_{dead})/(2\lambda+d_{CoFeB})$ [14], we obtain a volume saturation magnetization of $M_s = (643 \pm 21)$ emu/cm$^3$, and a magnetically dead layer thickness at each Nb/CoFeB interface of $d_{dead} = (0.33 \pm 0.09)$ nm, which is slightly thinner than those at Nb/Co ($d_{dead} = 0.4$ nm) and Nb/Fe ($d_{dead} = 0.55$ nm) interfaces [16]. Here, $\lambda = 110$ nm [17,18] is an estimate of the London penetration depth of polycrystalline Nb. $M_s$ obtained here is smaller than the maximum bulk magnetization of 1300 emu/cm$^3$ [10,19], implying a reduced magnetization in thin (< 5 nm) CoFeB and, possibly, partial oxidation or Ga implantation in nanopillars. For $d_{CoFeB} = 4.5$ nm, the magnetization of CoFeB switches at $\mu_0 H_c < M_s(d_{CoFeB} - d_{dead})/(2\lambda+d_{CoFeB})$ and hence the maximum in $I_c$ occurs at $\delta \approx \mu_0 H_c$, resulting in a spread in $\delta$ due to variations in $H_c$. A clear change in $\delta$ is not observed following thermal annealing, suggesting that the magnetization of CoFeB in nanopillars is unaffected by annealing, consistent with our magnetization measurements of unpatterned films (see supplementary Fig. S2). In the inset of Fig. 1(b), we have plotted the normalized magnetic field periodicity ($n$) of $I_c(H)$ versus $d_{CoFeB}$, where $n$ is determined from sinc ($n\Phi/\Phi_0$) with $\Phi = \mu_0 HL(2\lambda+d_{CoFeB})$, $L$ is the length of the junction perpendicular to the applied magnetic field, and $\Phi_0$ is the magnetic flux quantum. For the $d_{CoFeB}$ range investigated, $n \approx 1$, consistent with a dominant first harmonic current-phase relation.

In Fig. 2(a), we have plotted the total specific resistance of the nanopillars ($AR_N$) versus $d_{CoFeB}$ at 1.6 K before and after thermal annealing. From a least square regression line fit ($AR_N = \rho_{CoFeB} \times d_{CoFeB} + 2AR_{Nb/CoFeB}$), we estimate an as-grown CoFeB resistivity of $\rho_{CoFeB} = (88 \pm 46)$ µΩ·cm, which is higher than the resistivity of a Co$_{60}$Fe$_{40}$ polycrystalline ferromagnetic alloy ($\rho \approx 15$ µΩ·cm at 10 K [20]). We also estimate the specific resistances of the two Nb/CoFeB interfaces as $2AR_{Nb/CoFeB} = (4.4 \pm 1.4)$ fΩ·m$^2$. The effective electron mean free path in as-grown CoFeB is $l = 3\pi^2\hbar/k_F^2 e^2 \rho_{CoFeB} = (1.8 \pm 0.9)$ nm, where $\hbar$ is the Planck constant divided by $2\pi$, $k_F = 0.104$ nm$^{-1}$ [21] is the Fermi wave number in the majority band of CoFeB, and $e$ is the elementary charge. We observe a decrease in $AR_N$ for all the devices following thermal annealing, suggesting a decrease in $\rho_{CoFeB}$ (and increase in the electron mean free path) as a result of increased structural order.

The increased degree of scatter in $AR_N$ versus $d_{CoFeB}$ for the nanopillars after thermal annealing is likely due to the variation of the resistance of CoFeB and Nb/CoFeB interfaces induced by annealing.

The decrease in $R_N$ through thermal annealing results in a notable enhancement of the Josephson critical current density ($J_c$) as shown in Fig. 2(b). The relative $J_c$ change following thermal annealing [defined as ($J_{c,annealed} - J_{c,as-grown}$) / $J_{c,as-grown}$] is 80 – 700% depending on the CoFeB thickness [see inset of Fig. 2(b)]. Whilst an enhancement of $J_c$ has been observed for all nanopillars investigated, the relative $J_c$ change does not show a clear $d_{CoFeB}$ dependence due to the relatively large variation in $R_N$ induced by thermal annealing.

To investigate the effect of thermal annealing upon the proximity coherence length of superconductivity in CoFeB, in Fig. 3(c) we have plotted the characteristic voltage ($I_cR_N$) versus $d_{CoFeB}$ at 1.6 K before and after thermal annealing. In superconductor/ferromagnet/superconductor Josephson devices, $I_cR_N$ typically displays damped oscillatory behaviour as a function of ferromagnetic barrier thickness due to 0-π phase transitions [16,22–26]. In our devices, although $I_cR_N$ exponentially decays with $d_{CoFeB}$ we do not observe evidence of 0-π oscillations. The apparent absence of these oscillations is likely due to a strong magnetic exchange energy of CoFeB ($E_{ex} \approx k_B T_{Curie}$ = 113 meV where $k_B$ is the Boltzmann constant and $T_{Curie}$ = 1313 K [27] is the Curie temperature of CoFeB), giving rise to a short oscillation period; $\pi v_F \hbar / 2 E_{ex}$ [28] ≈ 0.1 nm where $v_F$ = 1.2×10$^4$ m/s [21] is the Fermi velocity in CoFeB. Hence, the oscillation is smoothed out by the thickness variation (roughness) of CoFeB and is undetectable. As shown in the inset of Fig. 2(c), the relative change of $I_cR_N$ through annealing [defined as ($I_cR_{N,annealed} - I_cR_{N,as-grown}$) / $I_cR_{N,as-grown}$] increases with increasing $d_{CoFeB}$ (i.e., the decay slope of $I_cR_N$ with $d_{CoFeB}$ becomes shallower as a result of annealing). By fitting the decay slope to $I_cR_N \propto \exp(-\xi_{CoFeB}/d_{CoFeB})$, the proximity coherence length of superconductivity in CoFeB ($\xi_{CoFeB}$) is estimated to be (2.15 ± 0.10) nm and (2.41 ± 0.12) nm for the devices before and after annealing, respectively. Considering the fact that the magnetic moment of CoFeB is unchanged following thermal annealing [see Fig. 1(b)], the enhancement of $\xi_{CoFeB}$ is likely due to the increase in the electron mean free path of CoFeB as a result of improved structural ordering, consistent with the decrease in $R_N$ in Fig. 2(a).

In conclusion, we have demonstrated Josephson coupling through CoFeB alloy and its optimization with thermal annealing in Nb/CoFeB/Nb nanopillars. We have found a notable enhancement of the Josephson critical current up to a maximum of 700% following thermal annealing at 400°C, which is attributed to an increase in the proximity coherence length of superconductivity in CoFeB and improved transparency at the Nb/CoFeB interfaces. The thermal optimization of Josephson coupling following thermal annealing is attractive for the development of energy efficient superconducting spintronic devices.

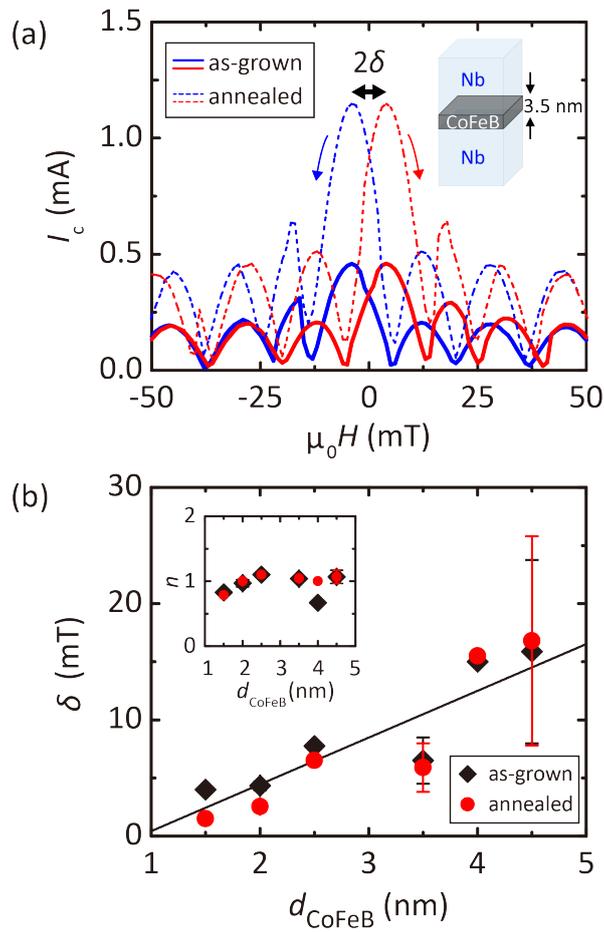

FIG. 1. (a) An $I_c(H)$ pattern for a Nb(300 nm)/CoFeB(3.5 nm)/Nb(300 nm) nanopillar before (solid lines) and after (dashed lines) thermal annealing at 400°C for 30 minutes. The red (solid/dashed) line shows $I_c$ with increasing $H$ and the blue (solid/dashed) line shows $I_c$ with decreasing $H$. (b) In-plane magnetic hysteresis ($\delta$) for Nb(300 nm)/CoFeB($d_{CoFeB}$)/Nb(300 nm) nanopillars before (black diamonds) and after (red circles) thermal annealing. The vertical error bars represent the statistical scatter of $\delta$ for multiple nanopillars measured on the same circuit. The black line is a least-squares regression line fit for the nanopillars before thermal annealing, giving a volume saturation magnetization of (643 ± 21) emu/cm$^3$ and a magnetically dead layer thickness at each Nb/CoFeB interface of (0.33 ± 0.09) nm. The inset shows the magnetic field periodicity ($n$) of $I_c(H)$ vs. $d_{CoFeB}$. All data at 1.6 K.

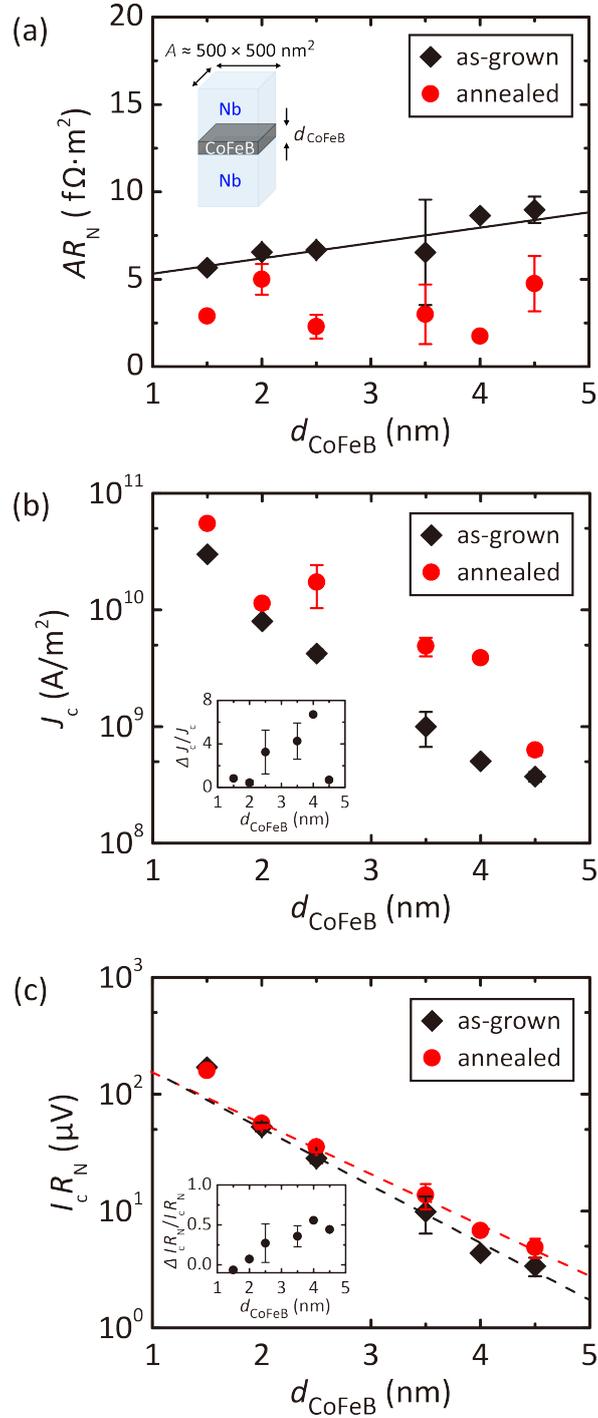

FIG. 2. (a) $AR_N$ vs. $d_{CoFeB}$ before (black diamonds) and after thermal annealing (red circles). The black line shows a least-squares regression line fit for the nanopillars before annealing from which we estimate $\rho_{CoFeB} \approx (88 \pm 46)$ μΩ·cm and $2AR_{Nb/CoFeB} = (4.4 \pm 1.4)$ fΩ·m². (b) $J_c$ vs. $d_{CoFeB}$ before (black diamonds) and after thermal annealing (red circles) at 400°C for 30 minutes. The inset shows the relative change in $J_c$ following thermal annealing $[(J_{c,annealed} - J_{c,as-grown}) / J_{c,as-grown}]$ vs. $d_{CoFeB}$. (c) $I_cR_N$ vs. $d_{CoFeB}$ where the dashed lines are least-square regression line fits giving a proximity coherence length in CoFeB of $\xi_{CoFeB} = (2.15 \pm 0.10)$ nm and $(2.41 \pm 0.12)$ nm, for the nanopillars before and after thermal annealing, respectively. The inset shows the relative change in $I_cR_N$ following thermal annealing $[(I_cR_{N,annealed} - I_cR_{N,as-grown}) / I_cR_{N,as-grown}]$ vs. $d_{CoFeB}$. The error bars in $AR_N$, $J_c$ and $I_cR_N$ represent the statistical scatter for multiple nanopillars. All data at 1.6 K.

**ACKNOWLEDGEMENTS**

The authors acknowledge funding from the EPSRC Programme Grants (No. EP/N017242/1 and No. EP/P026311/1). J.E.T. acknowledges funding from DTP EPSRC Grants (No. EP/M508007/1 and EP/N509620/1). J.W.A.R. acknowledges funding from the Royal Society through a University Research Fellowship.

**Supplementary information**
**for**
**Thermal annealing enhancement of Josephson critical currents in ferromagnetic CoFeB**

Sachio Komori*, Juliet E. Thompson*

Guang Yang, Graham Kimbell, Nadia Stelmashenko

Mark G. Blamire and Jason W. A. Robinson†

Department of Materials Science & Metallurgy, University of Cambridge,

27 Charles Babbage Road, Cambridge CB3 0FS, United Kingdom

*These authors contributed equally to this work.

†Corresponding author: jjr33@cam.ac.uk

## 1. Fraunhofer patterns of Nb(300 nm)/CoFeB(1.5 – 4.5 nm)/Nb(300 nm) nanopillars

In Fig. S1, we have plotted the critical current ($I_c$) vs. external in-plane magnetic field ($H$) for Nb(300 nm)/CoFeB(1.5 – 4.5 nm)/Nb(300 nm) nanopillars at 1.6 K before (solid lines) and after (dashed lines) thermal annealing at 400°C for 30 minutes in vacuum. All the data sets in the manuscript have been obtained from these 11 nanopillars prepared from a single sputtering deposition. The critical current density ($J_c$), the normal state resistance ($R_N$), the characteristic voltage ($I_c R_N$), and the total specific resistance ($AR_N$) of the corresponding nanopillars are summarized in Table S1.

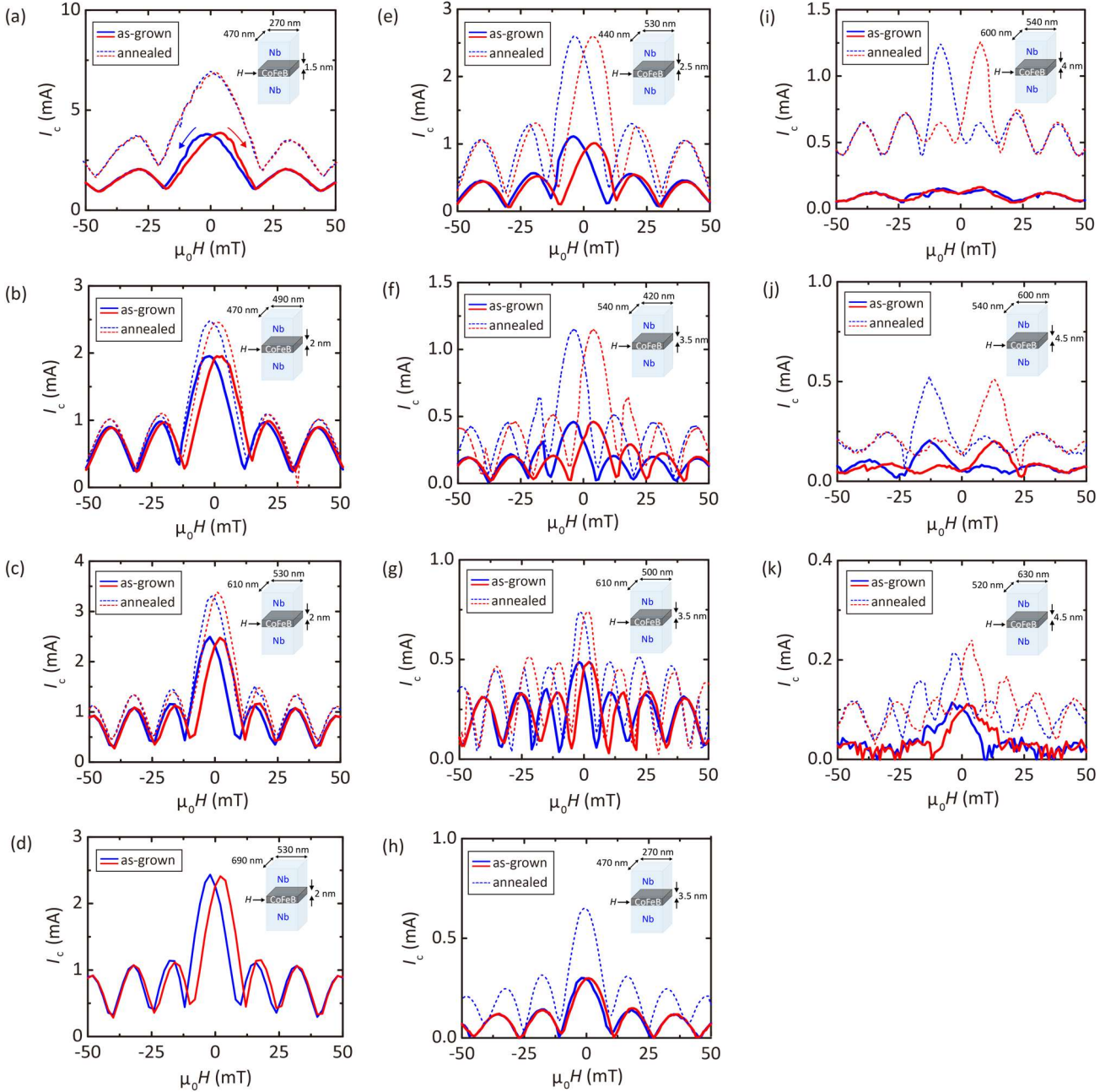

FIG. S1. $I_c(H)$ patterns for Nb(300 nm)/CoFeB(1.5 – 4.5 nm)/Nb(300 nm) nanopillars before (solid lines) and after (dashed lines) thermal annealing. The red lines show $I_c$ with increasing $H$ and the blue lines show $I_c$ with decreasing $H$. The inset shows the size of the nanopillars and the thickness of CoFeB. All data at 1.6 K.

**Table S1.** Critical current density ($J_c$), normal state resistance ($R_N$), characteristic voltage ($I_cR_N$), and total specific resistance ($AR_N$) of Nb(300 nm)/CoFeB(1.5 – 4.5 nm)/Nb(300 nm) devices at 1.6 K. The values after annealing are shown in brackets.

|   | $d_{CoFeB}$ (nm) | $J_c$ (A/m$^2$) | $R_N$ (mΩ) | $I_cR_N$ (μV) | $AR_N$ (fΩm$^2$) |
|---|---|---|---|---|---|
| a | 1.5 | $3.01 \times 10^{10}$ ($5.50 \times 10^{10}$) | 45.0 (23.0) | 170 (159) | 5.66 (2.89) |
| b | 2.0 | $8.50 \times 10^{9}$ ($1.07 \times 10^{10}$) | 29.4 (25.2) | 57.4 (62.0) | 6.75 (5.79) |
| c | 2.0 | $7.79 \times 10^{9}$ ($1.06 \times 10^{10}$) | 20.7 (16.2) | 51.6 (54.7) | 6.62 (5.17) |
| d | 2.0 | $7.69 \times 10^{9}$ ($1.29 \times 10^{10}$) | 17.3 (11.1) | 48.4 (52.1) | 6.30 (4.03) |
| e | 2.5 | $4.61 \times 10^{9}$ ($2.94 \times 10^{10}$) | 28.8 (12.2) | 30.9 (31.4) | 6.70 (2.84) |
| f | 3.5 | $1.74 \times 10^{8}$ ($4.35 \times 10^{9}$) | 21.3 (11.0) | 9.80 (12.6) | 5.63 (2.89) |
| g | 3.5 | $1.01 \times 10^{9}$ ($6.06 \times 10^{9}$) | 13.2 (4.44) | 6.43 (8.12) | 3.98 (1.34) |
| h | 3.5 | $1.33 \times 10^{9}$ ($4.28 \times 10^{9}$) | 44.3 (21.3) | 13.3 (20.3) | 10.0 (4.75) |
| i | 4.0 | $5.05 \times 10^{8}$ ($3.89 \times 10^{9}$) | 26.7 (5.39) | 4.36 (6.79) | 8.64 (1.75) |
| j | 4.5 | $4.09 \times 10^{8}$ ($6.95 \times 10^{9}$) | 33.3 (10.8) | 3.98 (5.78) | 9.73 (3.16) |
| k | 4.5 | $3.36 \times 10^{8}$ ($5.65 \times 10^{9}$) | 25.0 (19.2) | 2.76 (3.96) | 8.22 (6.33) |

## 2. Effect of thermal annealing on the magnetization of CoFeB

In Fig. S2, we have plotted the in-plane volume magnetization ($M$) versus $H$ for an unpatterned Nb(35 nm)/CoFeB(4 nm)/Nb(2 nm) control sample measured at 10 K before (black curve) and after (red curve) thermal annealing at 400°C for 30 minutes in vacuum. $M_s$ and $H_c$ are unchanged following thermal annealing, suggesting that the exchange energy of CoFeB is unaffected by the amorphous-crystalline transition. The relatively small $M_s$ ($\approx 643 \pm 21$ emu/cm$^3$) estimated from $\delta$ [see Fig. 1(b) in the main manuscript] compared with that obtained from the magnetization measurement of this unpatterned control sample ($M_s \approx 900$ emu/cm$^3$) may be due to the existence of partial oxidation or Ga implantation in nanopillars.

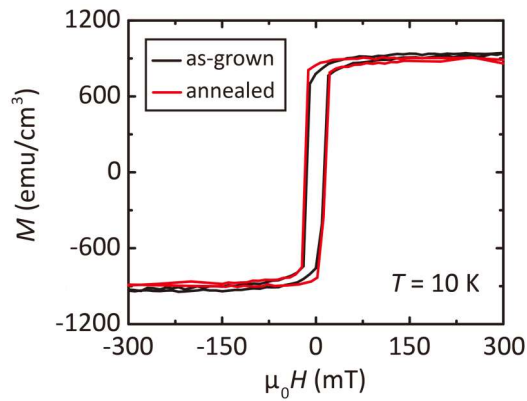

FIG. S2. $M$ ($H$) curve for an unpatterned Nb(35 nm)/CoFeB(4 nm)/Nb(2 nm) control sample measured at 10 K before (black curve) and after (red curve) thermal annealing.

### 3. Effect of thermal annealing on the superconductivity of Nb

To confirm that there is no significant effect of thermal annealing on the superconductivity of Nb, we prepared a Nb(30 nm) control sample and measured the superconducting transition temperature ($T_c$) before and after thermal annealing at 400°C for 30 minutes in vacuum. As shown in Fig. S3, $T_c$ is slightly suppressed (≈ 0.1 K) following thermal annealing, which is likely due to a slight oxidization of Nb as a result of a reaction with $SiO_2$ substrate. A slight $T_c$-suppression might result in a slight suppression of the Josephson critical currents, rather than the increase in Josephson critical currents observed in this study. Also, we note that the $T_c$ (= 9.2 K) of the 300-nm-thick Nb electrodes of all the devices used in this study is unchanged after thermal annealing. Hence, the annealing enhancement of critical currents observed in this study is due to the change in the electrical properties of CoFeB and CoFeB/Nb interfaces.

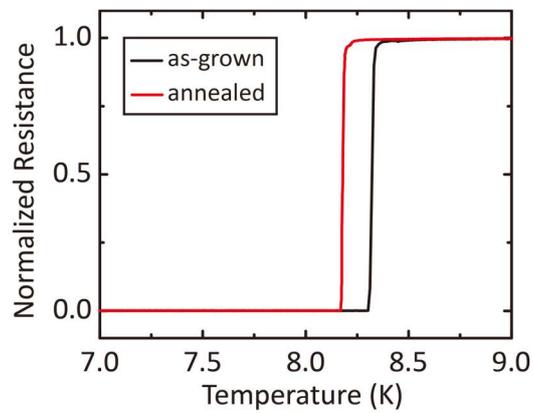

FIG. S3. Temperature dependence of the normalized resistance for a 30-nm-thick unpatterned Nb measured before (black curve) and after (red curve) thermal annealing at 400°C for 30 minutes in vacuum.